\documentclass[a4paper,10pt,twocolumn,twoside]{cpc-hepnp}
\usepackage{upgreek,fancyhdr}
\usepackage{multicol}
\usepackage{multirow}
\usepackage{graphicx}
\usepackage{booktabs}
\usepackage{amssymb,bm,mathrsfs,bbm,amscd}
\usepackage[tbtags]{amsmath}
\usepackage{lastpage}
\usepackage[english]{babel}
\usepackage{epsfig}
\usepackage{graphicx}
\usepackage{epstopdf}
\usepackage{graphbox}

\begin{document}

\fancyhead[c]{\small Chinese Physics C~~~Vol. xx, No. x (2024) xxxxxx}
\fancyfoot[C]{\small 010201-\thepage}
\footnotetext[0]{Received \today}

\title{The cosmic distance duality relation in light of the time-delayed strong gravitational lensing\thanks{Supported by the National Natural Science Fund of China under grant Nos. 12275034 and 12347101, and the Fundamental Research Funds for the Central Universities of China under grant No. 2024CDJXY-022.}}

\author{Li Tang$^{1,4}$
\quad Hai-Nan Lin$^{2,3;1)}$\email{linhn@cqu.edu.cn}
\quad Ying Wu$^{1,4}$}

\maketitle
\hspace{1cm}

\address{$^1$ Department of Math and Physics, Mianyang Teachers' College, Mianyang 621000, China\\
$^2$ Department of Physics, Chongqing University, Chongqing 401331, China\\
$^3$ Chongqing Key Laboratory for Strongly Coupled Physics, Chongqing University, Chongqing 401331, China\\
$^4$ Research Center of Computational Physics, Mianyang Teachers' College, Mianyang 621000, China}

\begin{abstract}
  The cosmic distance duality relation (DDR), which links the angular diameter distance and the luminosity distance, is a cornerstone in modern cosmology. Any deviati
  on from DDR may indicate new physics beyond the standard cosmological model. In this paper, we use four high-precision time-delayed strong gravitational lensing (SGL) systems provided by the H0LiCOW to test the validity of DDR. To this end, we directly compare the angular diameter distances from these SGL systems and the luminosity distances from the latest Pantheon+ compilation of SNe Ia. In order to reduce the statistical errors arising from redshift matching, the Gaussian process method is applied to reconstruct the distance-redshift relation from the Pantheon+ dataset. We parameterize the possible violation of DDR in three different models. It is found that all results confirm the validity of DDR at 1$\sigma$ confidence level. Additionally, Monte Carlo simulations based on the future LSST survey indicate that the precision of DDR could reach $10^{-2}$ level with 100 SGL systems.
\end{abstract}



\footnotetext[0]{\hspace*{-3mm}\raisebox{0.3ex}{$\scriptstyle\copyright$}2019
Chinese Physical Society and the Institute of High Energy Physics
of the Chinese Academy of Sciences and the Institute
of Modern Physics of the Chinese Academy of Sciences and IOP Publishing Ltd}%

\section{Introduction}\label{sec:introduction}

The cosmic distance duality relation (DDR), also known as the Etherington relation \cite{etherington1933lx}, plays a fundamental role in modern cosmology. This relation connects two critical measures of cosmic distance for an astronomical object: the luminosity distance $D_L$ and the angular diameter distance $D_A$. In a spacetime described by a metric theory of gravity, where the number of photons is conserved as they travel along null geodesics, the DDR can be formulated as $D_A(z)(1+z)^2/D_L(z)=1$, with $z$ representing the redshift of the astronomical object. Any deviation from the standard DDR may imply the presence of new physics. Various phenomena can lead to the violation of DDR, such as photons coupling with non-standard particles \cite{Bassett:2003vu}, dust extinction \cite{Corasaniti:2006cv}, variations in fundamental constants \cite{Ellis:2013cu}, systematic observational errors \cite{Holanda:2012ia}, and so on. Moreover, as a cornerstone relation in cosmology, the DDR has been extensively employed to probe gas mass density profiles \cite{Cao:2016ggq}, determine the shapes of galaxy clusters \cite{Holanda:2011hh}, and constrain cosmological parameters \cite{Kumar:2020ole,Liao:2020zko}. Therefore, verifying the validity of DDR is of paramount importance.

A variety of methodologies, both cosmological model-dependent and model-independent, are employed to probe the DDR. Model-dependent approaches evaluate the DDR under specific cosmological model assumptions \cite{DeBernardis:2006ii,Holanda:2010ay,Uzan:2004my,Avgoustidis:2010ju,Hu:2018yah}, whereas model-independent methods are generally plagued by large uncertainties \cite{Meng:2011nt,Li:2011exa,Santos-da-Costa:2015kmv,Liao:2015uzb,Zhou:2020moc}. Testing the DDR typically involves two distinct distance measures: luminosity distance and angular diameter distance. The most straightforward comparison entails evaluating these two distances at the same redshift. However, the simultaneous measurement of luminosity distance and angular diameter distance for a single astronomical object poses significant challenges, necessitating their derivation from disparate observations. Such observations may include Type Ia supernovae (SNe Ia), gamma-ray bursts (GRBs), $H(z)$ measurements, angular diameter distances of galaxy clusters, galaxy cluster gas mass fractions, Einstein radii of strong gravitational lenses (SGL), and gravitational waves, among others. To reconcile these measurements at a common redshift, several methodologies have been proposed, including the nearby SNe Ia method to match redshifts \cite{Li:2011exa,Liao:2015uzb}, Gaussian process (GP) techniques \cite{Kumar:2021djt,He:2022phb} or the artificial neural network (ANN) approaches \cite{Tang:2022ykd,Tonghua:2023hdz} for reconstructing the distance-redshift relation.

Among the various approaches for testing the DDR, SNe Ia have emerged as the most frequently utilized. For example, Li et al. \cite{Li:2011exa} employed the nearby SNe Ia method, combining the Union2 SNe Ia sample with two galaxy cluster samples. They assessed the DDR at 1$\sigma$ confidence level with 18 galaxy clusters, and at 2$\sigma$ confidence level with 38 galaxy clusters. Liao et al. \cite{Liao:2015uzb} combined the SGL systems with the JLA SNe Ia compilation, and confirmed the validity of DDR at 1$\sigma$ confidence level. Similarly, Zhou et al. \cite{Zhou:2020moc} verified the DDR using the Pantheon SNe Ia sample and SGL systems, achieving 1$\sigma$ confidence level. Employing the GP method to reconstruct the distance-redshift relation from the Pantheon dataset, Lin et al. \cite{Lin:2018qal} corroborated the validity of the DDR at 1$\sigma$ confidence level using two galaxy cluster samples. Liu et al. \cite{Tonghua:2023hdz} compared the precision of DDR tests using data reconstructed by GP and ANN methods, and revealed that the GP method provides higher precision (10$^{-3}$ level) compared to the ANN method (10$^{-2}$ level).

The integration of refined astrophysical datasets and forthcoming observational advancements is poised to enhance the precision of testing the DDR \cite{Liao:2019xug,Lin:2019mrl,Lin:2020vqj,Arjona:2020axn,Renzi:2020bvl,Huang:2024zvk}. Recently, Scolnic et al. \cite{Scolnic:2021amr} presented the Pantheon+ dataset, a comprehensive and meticulously curated compilation that substantially expands the catalog of SNe Ia light curves available for cosmological studies. The Pantheon+ dataset encompasses 1701 light curves from 1550 spectroscopically confirmed SNe Ia. The cross-calibration of photometric systems within the Pantheon+ sample, along with the subsequent re-calibration of the SALT2 light-curve model \cite{Brout:2021mpj}, significantly reduces the systematic uncertainties and enhances the accuracy of distance modulus measurements. Additionally, the $H_0$ Lenses in COSMOGRAIL's Wellspring (H0LiCOW) collaboration has provided six gravitationally lensed quasars with precisely measured time-delay distances \cite{Suyu:2009by,Suyu:2013kha,Suyu:2016qxx,Wong:2016dpo,Jee:2019hah,Chen:2019ejq,Birrer:2018vtm,Rusu:2019xrq}, constraining the Hubble constant $H_0$ to a precision of 2.4$\%$ within the standard flat $\Lambda$CDM model \cite{Wong:2019kwg}. From the time-delayed gravitational lensing, we can extract the information of angular diameter distances, thus offering a valuable opportunity to test the DDR.

In this study, we will test the validity of DDR using the combination of luminosity distances from the Pantheon+ dataset and angular diameter distances from SGL systems provided by the H0LiCOW. To obtain the luminosity distance at the redshifts of the SGL systems, we will employ the GP method to reconstruct the distance-redshift relation from the Pantheon+. The remainder of this paper is structured as follows: Section {\ref{sec:data}} details the data and methodology employed in testing the DDR. Specifically, Section {\ref{sec:data_DA}} illustrates how to measure angular diameter distance from SGL systems; Section {\ref{sec:data_DL}} illustrates the GP reconstruction of distance-redshift relation from the Pantheon+; Section {\ref{sec:DDR}} shows how to combine the SGL systems and SNe Ia to test DDR. In Section {\ref{sec:simulation}}, we use Monte Carlo simulations to predict the prospect of a larger sample of SGL data in constraining the DDR. Finally, Section {\ref{sec:conclusions}} offers a discussion of the findings and draws conclusions based on the analysis.

\section{Data and methodology}\label{sec:data}

\subsection{Angular diameter distance from SGL}\label{sec:data_DA}
Strong gravitational lensing is a potent tool for exploring the cosmos. When light from a luminous source, such as a quasar, passes near a massive object-the lens-it bends, resulting in observable images of the source. Due to different paths light can take, multiple images of the source can appear. The temporal offset between these images is known as the time delay, which can be written as \cite{Mollerach:2002}
\begin{equation}\label{eq:time-delay}
    \Delta t=\frac{D_{\Delta_t}\Delta\phi(\bm{\xi}_{\mathrm{lens}})}{c},
\end{equation}
where $c$ is the speed of light, $\Delta\phi$ represents the Fermat potential difference between the two images, and $\bm{\xi}_{\mathrm{lens}}$ denotes the parameters of the lens model. The time-delay distance $D_{\Delta_t}$ is defined by \begin{equation}\label{eq:D_Dt}
     D_{\Delta_t}=(1+z_l)\frac{D_A^l D_A^s}{D_A^{ls}},
\end{equation}
where $z_l$ is the redshift of the lens, and $D_A^l$, $D_A^s$ and $D_A^{ls}$ correspond to the angular diameter distances between the observer and the lens, between the observer and the source, and between the lens and the source, respectively. If the time delay is precisely measured and the Fermat potential is well modeled, we can determine the time-delay distance $D_{\Delta t}$ according to Eq.(\ref{eq:time-delay}).

By incorporating kinematic information from the lens galaxy-such as the light profile $\bm{\xi}_{\mathrm{light}}$, the projected stellar velocity dispersion $\sigma^P$, and the anisotropy distribution of the stellar orbits parameterized by $\beta_{\mathrm{ani}}$, the angular diameter distance from the observer to the lens can be determined by \cite{Birrer:2018vtm}
\begin{equation}\label{eq:D_A_l}
    D_A^l=\frac{1}{1+z_l}D_{\Delta t}\frac{c^2 \bm{J}(\bm{\xi}_{\mathrm{lens}},\bm{\xi}_{\mathrm{light}},\beta_{\mathrm{ani}})}{(\sigma^P)^2},
\end{equation}
where $\bm{J}$ is a function that encapsulates all the model components, derived from angular measurements on the sky and the distribution of stellar orbital anisotropy.

According to the Planck 2018 results \cite{Planck:2018vyg}, our universe is spatially flat. In such a flat universe, the angular diameter distance at redshift $z$ can be expressed in terms of the comoving distance $r(z)$ as $D_A=r(z)/(1+z)$. The comoving distance between the lens and the source is given by $r^{ls}=r(z_s)-r(z_l)$, where $z_l$ and $z_s$ are the redshifts of lens and source, respectively \cite{Bartelmann:1999yn}. Consequently, the angular diameter distance between the lens and the source can be written as
\begin{equation}\label{eq:D_A_ls}
    D_A^{ls}=D_A^s-\frac{1+z_l}{1+z_s}D_A^l.
\end{equation}

The COSmological MOnitoring of GRAvItational Lenses (COSMOGRAIL) program \cite{Eigenbrod:2005ie,Bonvin:2018dcc} is dedicated to the precise measurement of time delays in well-known lensed quasars, utilizing a network of dedicated telescopes located in both the northern and southern hemispheres. Leveraging these time delay measurements alongside detailed models of the mass distribution of the lensing galaxies and their environments, the H0LiCOW collaboration \cite{Suyu:2016qxx} aims to determine the Hubble constant $H_0$ with high precision. Recently, the H0LiCOW collaboration constrained the Hubble constant to an accuracy of 2.4$\%$ using a combined sample of six lenses \cite{Wong:2019kwg}. The posterior distributions for the time-delay distances $D_{\Delta_t}$ of these six lens systems are available on the H0LiCOW website\footnote{http://www.h0licow.org}. Additionally, the angular diameter distances $D_A^l$ for four of these six lenses, i.e. RXJ1131-1231, PG 1115+80, B1608+656, and SDSS 1206+4332, are also provided.

\begin{figure*}[htbp]
 \centering
 \includegraphics[width=0.5\textwidth]{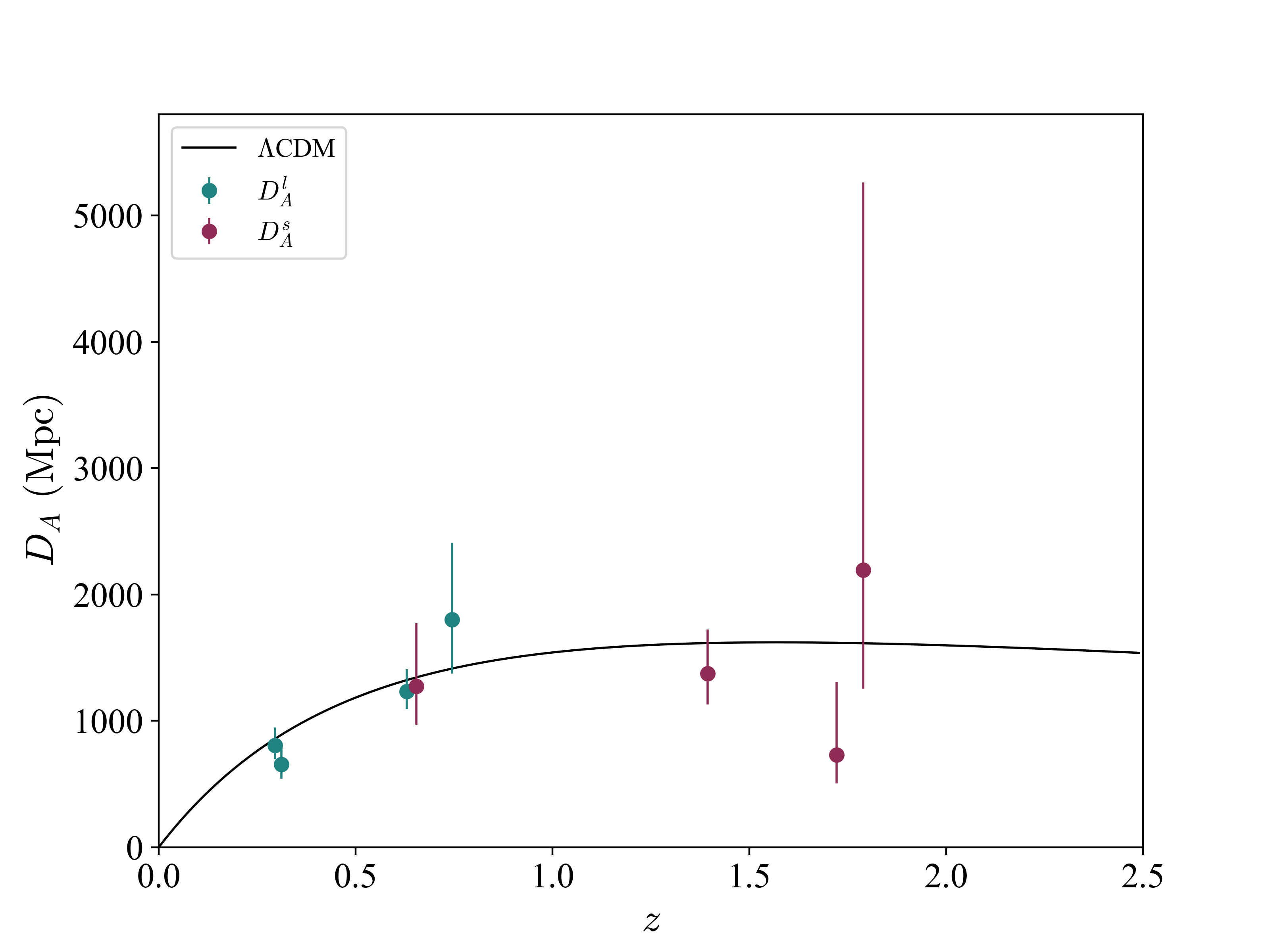}
 \caption{The angular diameter distance from four SGL systems. The black line represents the best-fitting curve of the $\Lambda$CDM model from the Pantheon+ dataset.}\label{fig:D_A}
\end{figure*}

In our work, we only consider the four SGL systems for which both the time-delay distance and angular diameter distance to the lens are available. The redshifts and distances of this sample are presented in Table \ref{tab:SGLsample}. Using the posterior distributions of $D_{\Delta_t}$ and $D_A^l$, we can determine the angular diameter distance to the source $D_A^s$ by combining Eq.(\ref{eq:D_Dt}) and Eq.(\ref{eq:D_A_ls}). For each system, we randomly sample $D_{\Delta_t}$ and $D_A^l$ from their respective posterior distributions 1000 times, and then compute the distribution of $D_A^s$. Consequently, our dataset comprises a total of eight angular diameter distances (one $D_A^l$ and one $D_A^s$ for each of the four SGL systems). To illustrate these results, we plot the angular diameter distances with 1$\sigma$ uncertainty in Figure \ref{fig:D_A}. The best-fitting $\Lambda$CDM model curve derived from the Pantheon+ dataset is also over-plotted for comparison. As is seen, most data points are consistent with the $\Lambda$CDM curve, regardless of the large uncertainty for some data points. Note that the data points are cosmological model-independent, except the spatially flat assumption.

\begin{table}[htbp]
\centering
\caption{\small{Redshifts and distances of the SGL sample ordered by lens redshift.}}\label{tab:SGLsample}
\arrayrulewidth=1.0pt
\renewcommand{\arraystretch}{1.3}
{\begin{tabular}{cccccc} 
\hline\hline 
Lens  & $z_l$& $z_s$ & $D_{\Delta_t}$ (Mpc) & $D_A^l$ (Mpc) & Reference\\\hline
RXJ1131$-$1231& 0.295& 0.654& $2096_{-83}^{+98}$& $804_{-112}^{+141}$& \cite{Chen:2019ejq,Suyu:2013kha}\\
PG 1115+080& 0.311& 1.722& $1470_{-127}^{+137}$& $697_{-144}^{+186}$& \cite{Chen:2019ejq}\\
B1608+656& 0.630& 1.394& $5156_{-236}^{+296}$& $1228_{-151}^{+177}$& \cite{Suyu:2009by,Jee:2019hah}\\
SDSS 1206+4332& 0.745& 1.789& $5769_{-471}^{+589}$& $1805_{-398}^{+555}$& \cite{Birrer:2018vtm}\\
\hline
\end{tabular}}
\end{table}

\subsection{Luminosity distance from SNe Ia}\label{sec:data_DL}

SNe Ia are invaluable as distance indicators in cosmological research due to their nearly constant intrinsic luminosity. The latest Pantheon+ compilation, detailed in Ref.\cite{Scolnic:2021amr}, integrates data from 18 distinct surveys, yielding 1701 light curves from 1550 spectroscopically confirmed SNe Ia across a redshift range of $z\in[0.001,2.3]$. According to the modified Tripp relation \cite{Tripp:1997wt}, the observed distance modulus of SNe Ia at redshift $z$ is expressed as
\begin{equation}\label{eq:mu}
    \mu=m_B-M+\alpha x_1-\beta c_1-\delta_{\mu-\mathrm{bias}},
\end{equation}
where $m_B$ is the apparent magnitude in the B-band, and $M$ is the absolute magnitude. The parameters $x_1$ and $c_1$ represent the stretch and color of the supernovae, respectively. The term $\delta_{\mu-\mathrm{bias}}$ accounts for the bias correction related to selection effects and other simulation-related issues. Using the BEAMS with bias corrections (BBC) method to calibrate SNe Ia \cite{Kessler:2016uwi}, the corrected magnitude $m_{\mathrm{corr}}$, and the corresponding covariance matrix are made available\footnote{https://pantheonplussh0es.github.io/}. Consequently, Eq.(\ref{eq:mu}) can be simplified to
\begin{equation}\label{eq:mu_corr}
    \mu=m_{\mathrm{corr}}-M.
\end{equation}
We convert the distance modulus to the luminosity distance using the relation
\begin{equation}
    \mu=5\log_{10}\frac{D_L(z)}{\mathrm{Mpc}}+25.
\end{equation}

To determine the luminosity distance at any redshift, we employ the GP method to reconstruct the distance-redshift relation from the Pantheon+ dataset without any parametric assumptions. The GP method is widely used in cosmological research for its efficacy in handling regression and classification tasks. To reconstruct a function $y=f(x)$ from observational data $(x_i,y_i)$, the GP method assumes that the data points are generated from a joint Gaussian distribution:
\begin{equation}
    \bm y\sim \mathcal{N}(\bm{\mu},\textbf{K}(\bm x,\bm x)+\textbf{C}),
\end{equation}
where $\bm x=\{x_i\}$, $\bm y=\{y_i\}$ are the observational vectors, $\bm\mu$ is the means of the Gaussian distribution, $\textbf{C}$ is the covariance matrix of the data, $[\textbf{K}(\bm x,\bm x)]_{ij} =k(x_i,x_j)$ is another covariance matrix, known as the kernel, which encodes assumptions about the smoothness, periodicity, and other properties of the reconstructed function. In this work, the kernel is parameterized by a squared-exponential covariance function:
\begin{equation}
    k(x_i,x_j)=\sigma^2_f\exp \left[-\frac{(x_i-x_j)^2}{2l^2}\right],
\end{equation}
where $\sigma_f$ and $l$ are the hyperparameters which determine the amplitude of the random fluctuations and the coherence length of the fluctuation, respectively. These hyperparameters are optimized by maximizing the marginalized likelihood.

\begin{figure*}[htbp]
 \centering
 \includegraphics[width=0.5\textwidth]{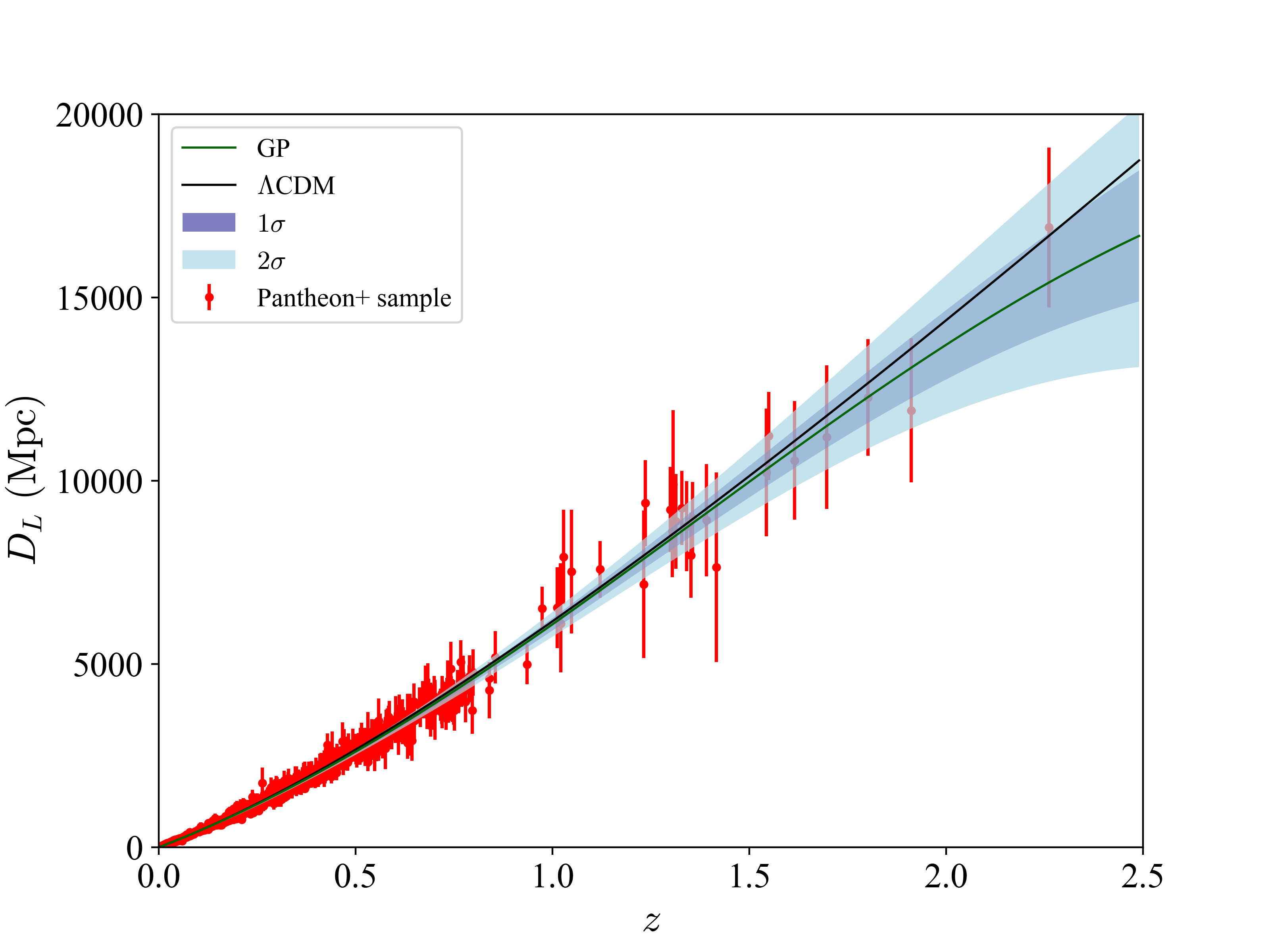}
 \caption{The $D_L-z$ relation reconstructed through the Gaussian process from the Pantheon+ at redshift $z<2.5$. The green curve is the central value of the reconstruction, and the blue regions are the $1\sigma$ and $2\sigma$ uncertainties of the reconstruction. The red dots with $1\sigma$ error bars are the Pantheon+ data points, and the black curve represents the best-fitting curve of the $\Lambda$CDM model.}\label{fig:D_L}
\end{figure*}

Using the luminosity distance data points obtained from Pantheon+, we perform the GP regression with the publicly available python package Gaussian Processes from Scikit-learn \cite{scikit-learn}. The reconstructed $D_L-z$ curve is plotted in Figure \ref{fig:D_L}. For comparison, we also plot the best-fitting curve of the $\Lambda$CDM model from the Pantheon+ dataset \cite{Brout:2022vxf}. As can be seen, the cosmological model-independent GP reconstruction is consistent with the best-fitting $\Lambda$CDM curve within 1$\sigma$ confidence level.

\subsection{The cosmic distance duality relation}\label{sec:DDR}

In the standard cosmological model, the relationship between angular diameter distance and luminosity distance at redshift $z$ is given by
\begin{equation}
    \frac{D_A(z)(1+z)^2}{D_L(z)}=1,
\end{equation}
which is known as the cosmic DDR. To test the potential violation of the standard DDR, we rewrite it as
\begin{equation}\label{eq:DDR}
    \frac{D_A(z)(1+z)^2}{D_L(z)}=\eta(z).
\end{equation}
The standard DDR holds when $\eta \equiv 1$. Deviation of $\eta$ from unity implies the violation of DDR. The parameter $\eta$ may be redshift-dependent, and various forms of $\eta$ have been proposed. In this work, we analyze the violation of DDR using three parameterized models:
\begin{align}
    &\textrm{Model~0}: ~\eta(z)=1+\eta_0,\\
    &\textrm{Model~1}: ~\eta(z)=1+\eta_0 z,\\
    &\textrm{Model~2}: ~\eta(z)=1+\eta_0 z/({1+z}),
\end{align}
where the DDR violation parameter $\eta_0$ is assumed to be a constant. The latter two models are designed to explore the possible redshift-dependence of DDR violation.

By combining the angular diameter distances from SGL with the corresponding luminosity distances from the GP reconstruction, we can constrain the DDR violation parameter in a cosmological model-independent manner. But be in mind that both the $D_A$ data points from SGL systems and the $D_L$ data points from SNe Ia are given in the form of probability distributions. The main processes are summarized as follows:
\begin{enumerate}
    \setlength{\itemsep}{-2pt}
    \item{For each redshift (four $z_l$ and four $z_s$) in the SGL sample, we randomly draw 1000 values of $D_A$ from the corresponding probability distribution presented in Section {\ref{sec:data_DA}}, and randomly draw 1000 values of $D_L$ from the corresponding probability distribution presented in Section {\ref{sec:data_DL}}, resulting in 1000 realizations at each given redshift.}
    \item{From the 1000 realizations at each given reshift, we calculate 1000 values of $\eta(z)$ according to Eq.(\ref{eq:DDR}), thus resulting the probability distribution of $\eta(z)$ at each redshift, denoted by $p_i(\eta(z_i))$.}
    \item{For a given model, such as Model 1, the joint likelihood can be written as $\mathcal{L}\propto\displaystyle\prod_{i=1}^{n}p_i(1+\eta_0z_i)$, where $n=8$ for the combination of four SGL systems, and $n=2$ for individual SGL system. For a given redshift $z$, $p_i(1+\eta_0z_i)=p_i(\eta(z_i))$, where $p_i(\eta(z_i))$ is obtained in step 2.}
    \item{The best-fitting parameter $\eta_0$ for each parameterized model is determined by maximizing the corresponding joint likelihood function.}
\end{enumerate}

\begin{figure*}[htbp]
 \centering
 \includegraphics[width=0.32\textwidth]{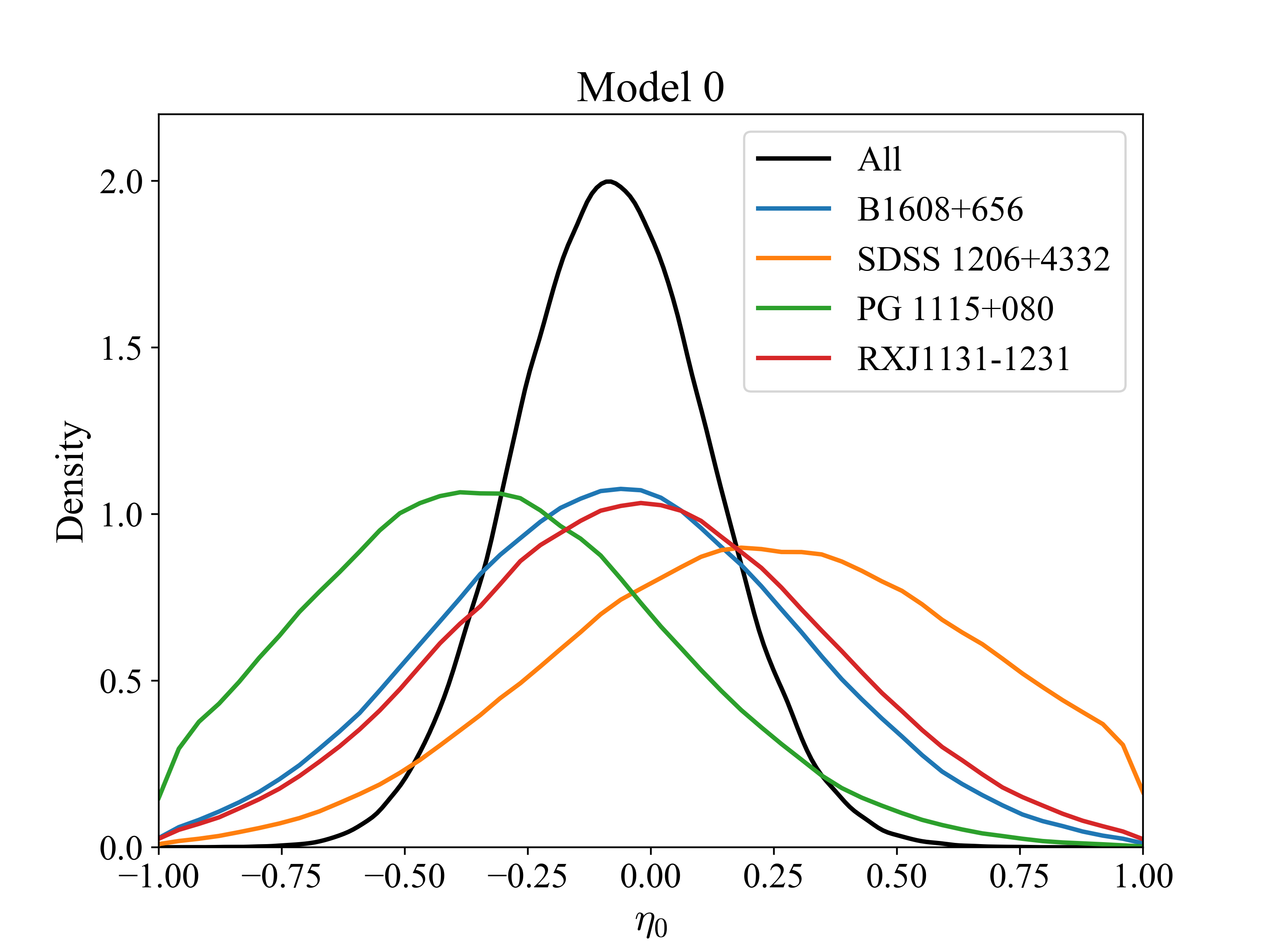}
 \includegraphics[width=0.32\textwidth]{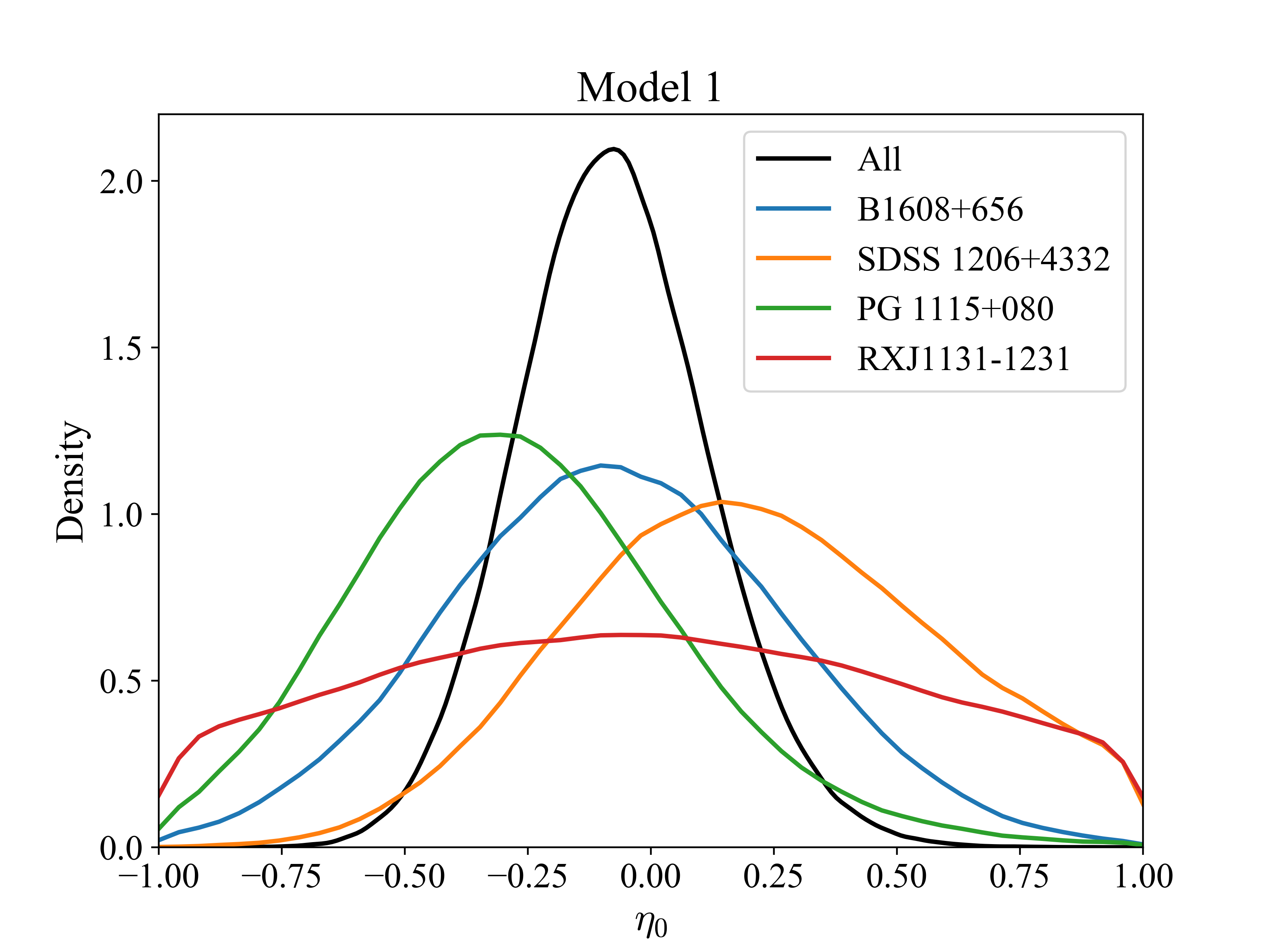}
 \includegraphics[width=0.32\textwidth]{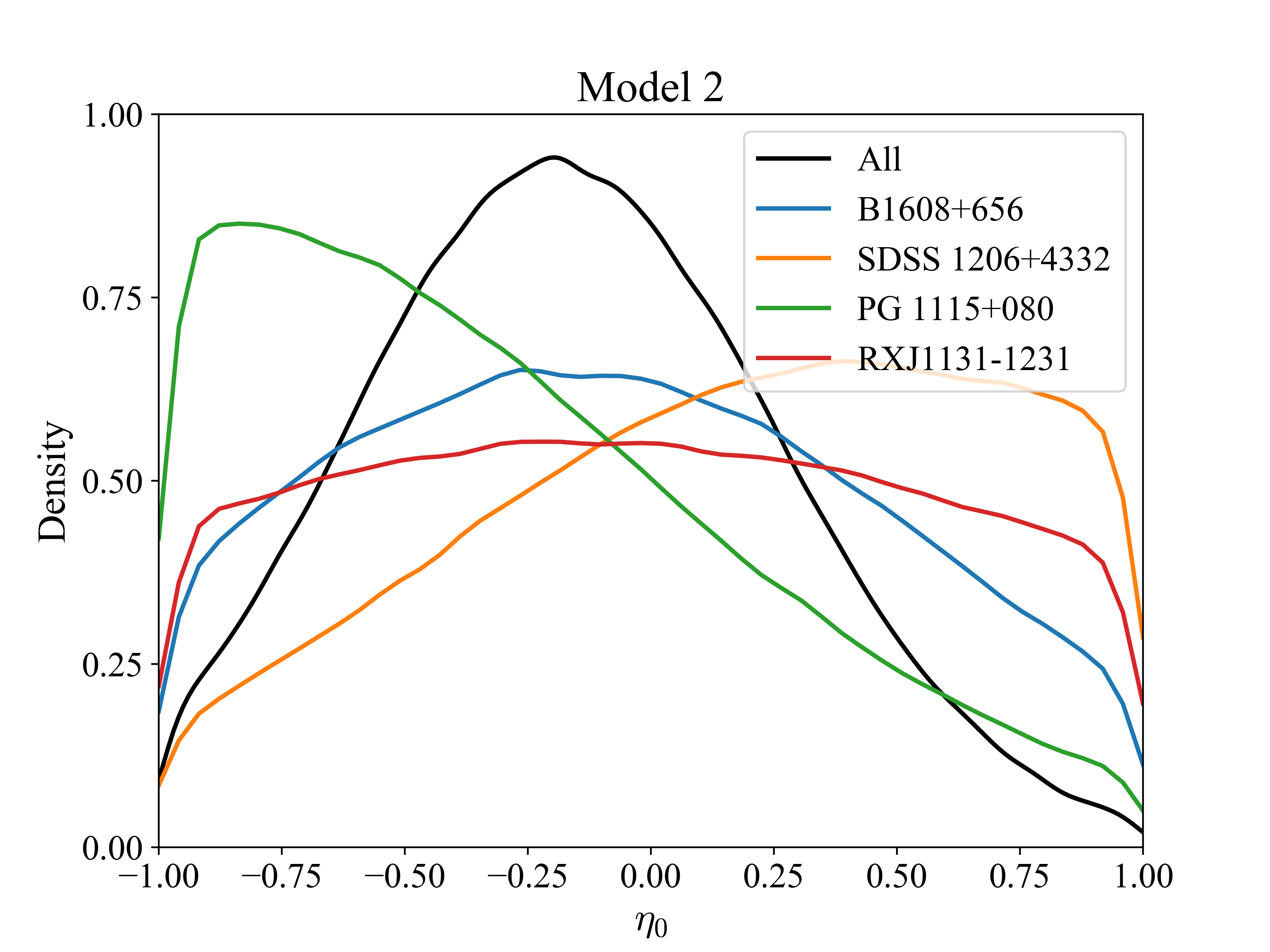}
 \caption{The posterior PDFs of $\eta_0$ in three types of parameterized models, constrained from the combination of all four SGL systems, as well as from individual SGL system.}\label{fig:eta0}
\end{figure*}

Assuming a flat prior on $\eta_0$ and limiting $-1\leq\eta_0\leq1$, we perform a Markov Chain Monte Carlo (MCMC) analysis to calculate the posterior probability density function (PDF) of the parameter space. The best-fitting parameters for the three parameterized models are presented in Table \ref{tab:parameters}, with all SGL systems as well as each individual SGL system. The corresponding PDFs are presented in Figure \ref{fig:eta0}. In the framework of Model 0, the parameter is constrained with $\eta_0=-0.078_{-0.196}^{+0.200}$ for all SGL systems, indicating that the DDR is valid at 1$\sigma$ confidence level. For the other two redshift-dependent models, there is no strong evidence for the redshift-evolution of DDR. The validity of DDR is verified at 1$\sigma$ confidence level, with the constraints $\eta_0=-0.081_{-0.186}^{+0.195}$ for Model 1 and $\eta_0=-0.166_{-0.405}^{+0.433}$ for Model 2, respectively.

\begin{table}[htbp]
\centering
\caption{\small{The best-fitting parameter $\eta_0$ in three types of parameterized models.}}\label{tab:parameters}
\arrayrulewidth=1.0pt
\renewcommand{\arraystretch}{1.3}
{\begin{tabular}{cccccc} 
\hline\hline 
Lens  & All& RXJ1131$-$1231 & PG 1115+080 & B1608+656 & SDSS 1206+4332\\\hline
Model 0& $-0.078_{-0.196}^{+0.200}$& $-0.021_{-0.382}^{+0.386}$& $-0.329_{-0.351}^{+0.380}$& $-0.067_{-0.367}^{+0.371}$& $0.227_{-0.433}^{+0.425}$\\
Model 1& $-0.081_{-0.186}^{+0.195}$& $-0.022_{-0.575}^{+0.596}$& $-0.286_{-0.306}^{+0.336}$& $-0.076_{-0.340}^{+0.351}$& $0.204_{-0.361}^{+0.403}$\\
Model 2& $-0.166_{-0.405}^{+0.433}$& $-0.029_{-0.631}^{+0.655}$& $-0.386_{-0.425}^{+0.623}$& $-0.090_{-0.553}^{+0.599}$& $0.210_{-0.628}^{+0.523}$\\
\hline
\end{tabular}}
\end{table}

Additionally, we investigate the DDR with each SGL system to elucidate their individual contributions to the overall constraint. The results are summarized in Table \ref{tab:parameters}. The corresponding PDFs are plotted in Figure \ref{fig:eta0}. Within the framework of each parameterized model, the constraint results for each of the four lensing systems consistently indicate the validity of the DDR at 1$\sigma$ confidence level. Notably, for the first three systems, the best-fitting $\eta_0$ prefers a negative value, consistent with the result constrained using all SGL systems. For the SDSS 1206+4332 system, however, the best-fitting $\eta_0$ prefers a positive value, although it is still consistent with zero due to the large uncertainty. This is because the angular diameter distances from this SGL system trend to be higher than the prediction of the $\Lambda$CDM model (see Figure \ref{fig:D_A}), while the corresponding luminosity distances are consistent with the $\Lambda$CDM prediction within 1$\sigma$ confidence level. Although the validity of the DDR is verified at 1$\sigma$ confidence level with current four SGL systems, the precision is only at $10^{-1}$ level.

\section{Monte Carlo simulations}\label{sec:simulation}

Although hundreds of SGL systems have been observed so far, only a tiny of them have time delay measurement. Leveraging realistic distributions for the properties of lenses and sources, and incorporating magnification bias and the detectability of image configurations, Oguri $\&$ Marshall \cite{Oguri:2010ns} conducted a comprehensive analysis of the expected yields for several planned surveys. Their results suggest that the forthcoming wide-field synoptic surveys are likely to identify several thousand lensed quasars in the next few years. Especially, the Large Synoptic Survey Telescope (LSST) is expected to discover over 8000 lensed quasars, and approximately 3000 of them are anticipated to have accurately measured time delays. Therefore, we employ Monte Carlo simulations to explore the precision of future datasets in constraining the DDR.

To simulate the SGL dataset $\{D_A^l, D_A^s\}$ and the corresponding uncertainties, we assume a fiducial cosmological model to determine the time-delay distance. In the fiducial $\Lambda$CDM model, the dimensionless comoving distance from $z_l$ to $z_s$ is given by
\begin{equation}\label{eq:r_LCDM}
    \Tilde{r}(z_l,z_s)=\int^{z_s}_{z_l}\frac{dz}{E(z)},
\end{equation}
where $E(z)=\sqrt{\Omega_m(1+z)^3+1-\Omega_m}$ is the dimensionless Hubble parameter, and $\Omega_m$ represents the normalized matter density. The angular diameter distance is expressed as
\begin{equation}\label{eq:DA_LCDM}
    D_A^{ls}=\frac{1}{1+z_s}\frac{c}{H_0}\Tilde{r}(z_l,z_s),
\end{equation}
Combining Eq.(\ref{eq:D_Dt}) and Eq.(\ref{eq:DA_LCDM}), the time-delay distance $D_{\Delta_t}$ can be calculated. From Eqs.(\ref{eq:D_Dt}) and (\ref{eq:D_A_ls}), the angular diameter distances of lens and source can be derived as
\begin{equation}\label{eq:D_A_l_sim}
    D_A^l=\frac{D_{\Delta_t} R_A}{1+z_l},
\end{equation}
\begin{equation}\label{eq:D_A_s_sim}
    D_A^s=\frac{1+z_l}{1+z_s}\frac{D_{\Delta_t} R_A}{1-R_A},
\end{equation}
where the distance ratio $R_A$ is defined by
\begin{equation}\label{eq:RA}
    R_A\equiv \frac{D_A^{ls}}{D_A^s}=\frac{c^2\theta_E}{4\pi\sigma^2_{\mathrm {SIS}}},
\end{equation}
with $\theta_E$ being the Einstein radius and $\sigma_{\mathrm{SIS}}$ the velocity dispersion of the lens galaxy modeled using the singular isothermal sphere (SIS) model. Here, the widely used and well-established SIS model offers a simple and analytically tractable framework to simulate and study the constraints on the DDR.

The uncertainties of $D_A^l$ and $D_A^s$ are propagated from that of $R_A$ and $D_{\Delta_t}$ as follows:
\begin{equation}\label{eq:D_A_l_err}
    \frac{\delta D_A^l}{D_A^l}=\frac{1}{1+z_l}\sqrt{\left(\frac{\delta D_{\Delta_t}}{D_{\Delta_t}}\right)^2+\left(\frac{\delta R_A}{R_A}\right)^2},
\end{equation}
\begin{equation}\label{eq:D_A_s_err}
    \frac{\delta D_A^s}{D_A^s}=\sqrt{\left(\frac{\delta D_{\Delta_t}}{D_{\Delta_t}}\right)^2+\left(\frac{\delta R_A}{R_A(1-R_A)}\right)^2}.
\end{equation}
The uncertainty in $D_{\Delta_t}$ propagates from the uncertainties in $\Delta t$ and $\Delta\phi$, while the uncertainty in $R_A$ propagates from the uncertainties in $\theta_E$ and $\sigma_{\mathrm{SIS}}$, expressed as follows:
\begin{equation}
    \frac{\delta D_{\Delta t}}{D_{\Delta t}}=\sqrt{\left(\frac{\delta \Delta t}{\Delta t}\right)^2+\left(\frac{\delta \Delta\phi}{\Delta\phi}\right)^2},
\end{equation}

\begin{equation}
    \frac{\delta R_A}{R_A}=\sqrt{\left(\frac{\delta \theta_E}{\theta_E}\right)^2+4\left(\frac{\delta \sigma_{\mathrm{SIS}}}{\sigma_{\mathrm{SIS}}}\right)^2}.
\end{equation}
According to the analysis by Liao et al. \cite{Liao:2017ioi}, the measurement accuracy of $\Delta t$ and $\Delta\phi$ in lensed quasars can reach 3$\%$. The accuracy of $\theta_E$ and $\sigma_{\mathrm{SIS}}$ are expected to be about 1$\%$ and 5$\%$ respectively, in the future LSST survey \cite{Cao:2019kgn}.

\begin{figure*}[htbp]
 \centering \includegraphics[width=0.45\textwidth]{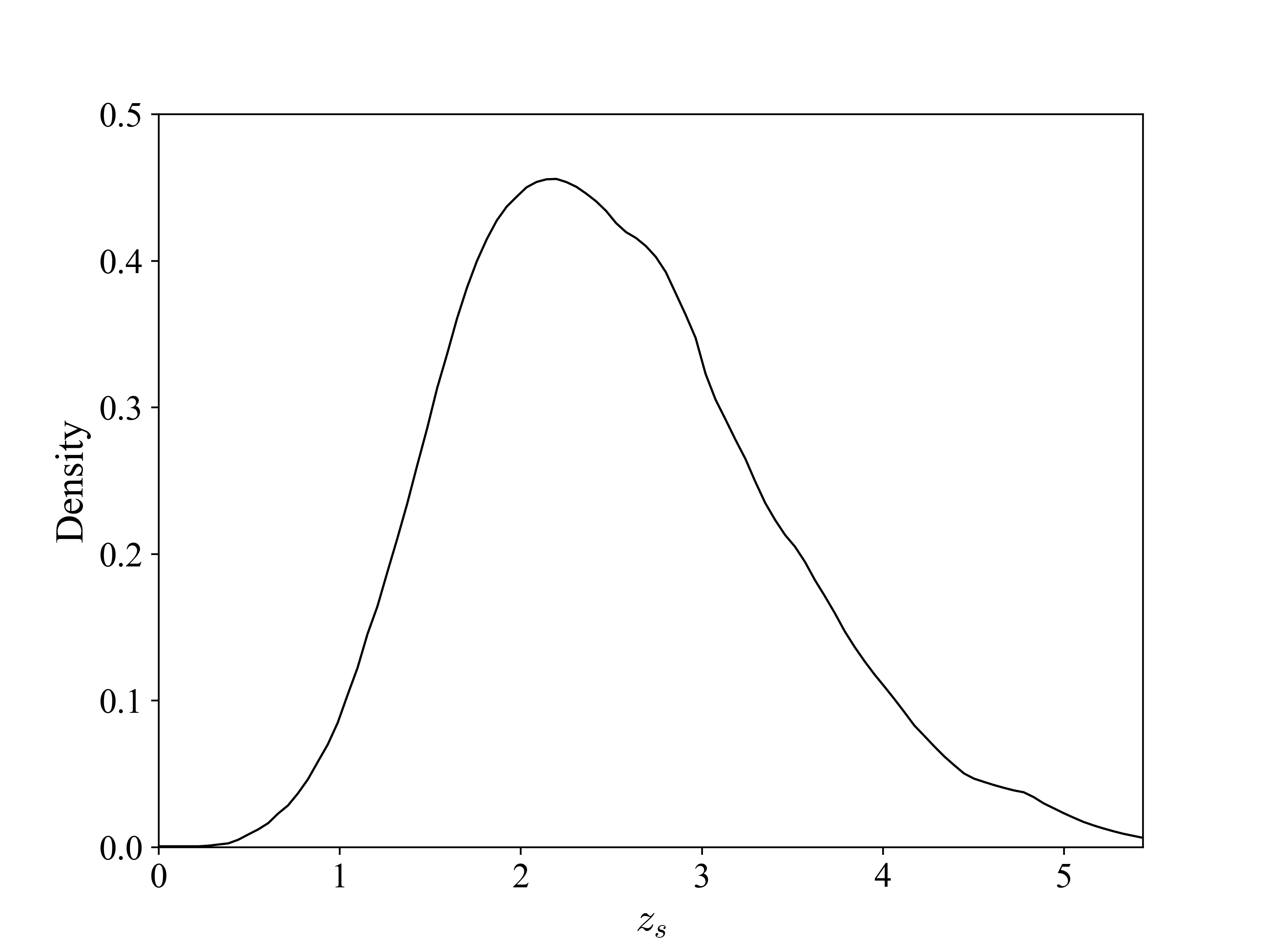}
 \caption{The distribution of $z_s$ of lensed quasars as is expected in the future LSST survey. Figure reproduced from Ref.\cite{Oguri:2010ns}.}\label{fig:z_s}
\end{figure*}

For the redshift distribution of SGL systems, Oguri $\&$ Marshall \cite{Oguri:2010ns} conducted a detailed analysis of the source redshift distribution of lensed quasars theoretically detectable by the LSST, presenting their findings in Figure 5 of Ref.\cite{Oguri:2010ns}, reproduced here in Figure \ref{fig:z_s}. For a quasar at redshift $z_s$, the probability density function of the lens redshift at $z_l$ is given by \cite{Biesiada:2014kwa}
\begin{equation}\label{eq:z_l_pdf}
    P(z_l|z_s)=C\frac{\Tilde{r}^2(z_l,z_s)\Tilde{r}^2(0,z_l)}{\Tilde{r}^2(0,z_s)E(z_l)},
\end{equation}
where $C$ is the normalization constant. Thus, given the redshift of the source, the lens redshift for a SGL system is randomly sampled according to this distribution. It should be mentioned that the redshift of simulated SGL systems is restricted to below 2.3, due to the restriction from the maximum redshift of Pantheon+ SNe. Although the combination of high-redshift data, such as gamma-ray bursts or gravitational waves can extend beyond the redshift limit of the SNe dataset, the large uncertainties in their calibration would introduce additional complexities and potential systematic errors. The availability and reliability of SNe as standard candles in the redshift range $z<2.3$, can minimize systematic uncertainties associated with the calibration and standardization of $D_L$ measurements. Therefore, the source redshift in our simulation is restricted to below 2.3.

Assuming the uncertainties for the parameters $(\Delta t,\Delta\phi, \theta_E,\sigma_{\mathrm{SIS}})$ are $(3\%,3\%,1\%,5\%)$, we can simulate the angular diameter distances of SGL system based on the fiducial $\Lambda$CDM model, where the cosmological parameters are fixed to be the best-fitting parameter from the Pantheon+ ($\Omega_m=0.334$, $H_0=73.6~{\rm km~s^{-1}~{Mpc}^{-1}}$) \cite{Brout:2022vxf}. The procedures are as follows:
\begin{enumerate}
    \setlength{\itemsep}{-2pt}
    \item Randomly sample the redshift $z_s$ from the source redshift distribution of lensed quasars presented in Figure \ref{fig:z_s}. If $z_s>2.3$, resample it again; otherwise continue.
    \item Given the source redshift $z_s$, sample the lens redshift $z_l$ according to the distribution given in Eq.(\ref{eq:z_l_pdf}).
    \item Calculate $D_{\Delta_t}$ and $R_A$ at $z_s$ and $z_l$, by combining Eqs.(\ref{eq:D_Dt}), (\ref{eq:r_LCDM}), (\ref{eq:DA_LCDM}) and (\ref{eq:RA}). The corresponding uncertainties $\delta D_{\Delta t}$ and $\delta R_A$ are determined with the assumed parameters accuracies.
    \item Calculate the fiducial angular diameter distances $\bar{D}_A^l$ and $\bar{D}_A^s$ according to Eqs.(\ref{eq:D_A_l_sim}) and $(\ref{eq:D_A_s_sim})$, respectively. The corresponding uncertainties $\delta D_A^l$ and $\delta D_A^s$ are determined using Eqs.(\ref{eq:D_A_l_err}) and (\ref{eq:D_A_s_err}).
    \item Sample $D_A^l$ and $D_A^s$ from the Gaussian distributions $D_A^l\sim \mathcal{N}(\bar{D}_A^l,\delta D_A^l)$ and $D_A^s\sim \mathcal{N}(\bar{D}_A^s,\delta D_A^s)$. Save the parameter set $(z_s,z_l,D_A^s, \delta D_A^s, D_A^l, \delta D_A^l)$ as an effective SGL system.
\end{enumerate}

\begin{figure*}[htbp]
 \centering
 \includegraphics[width=0.48\textwidth]{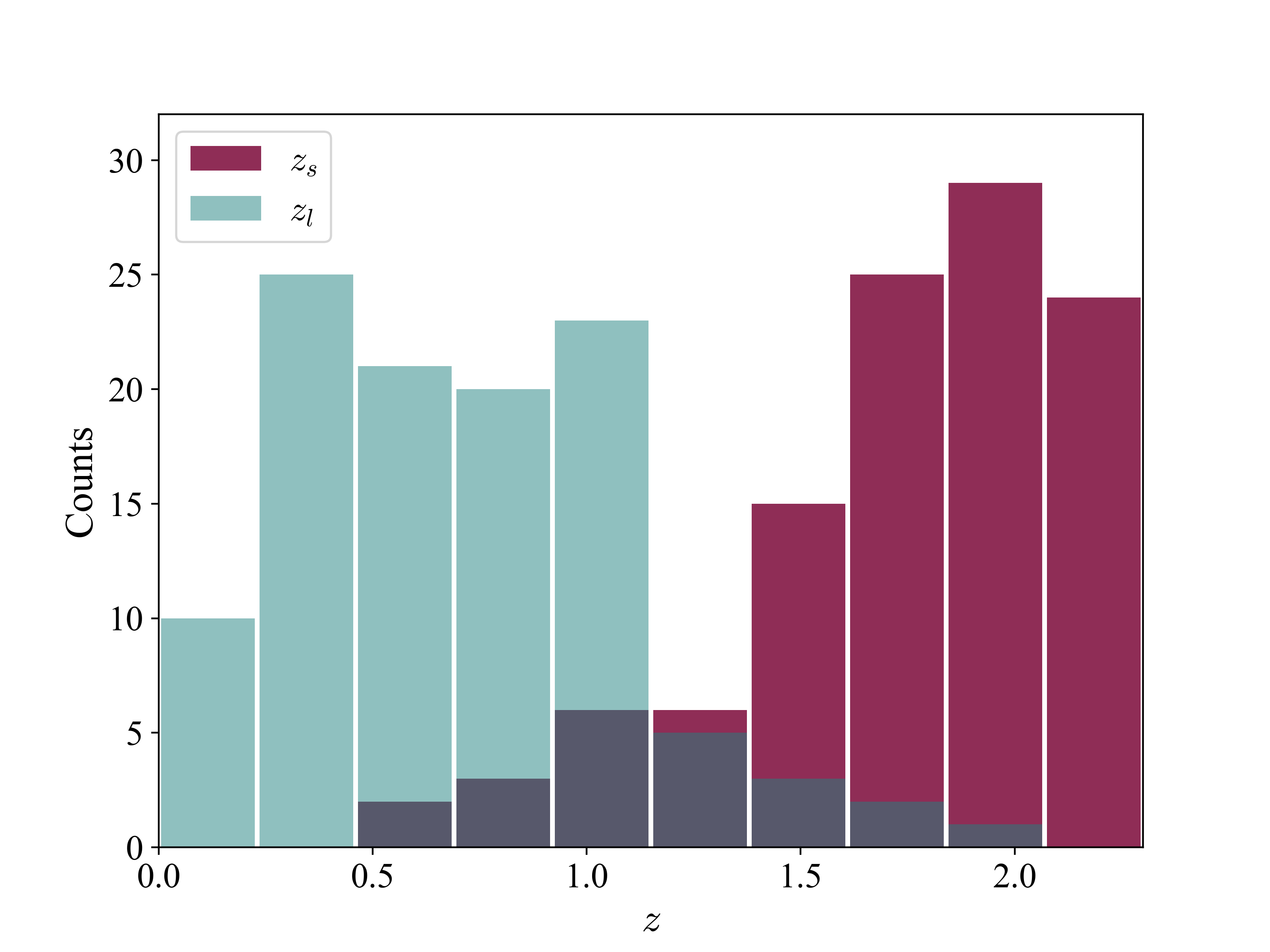}
 \includegraphics[width=0.48\textwidth]{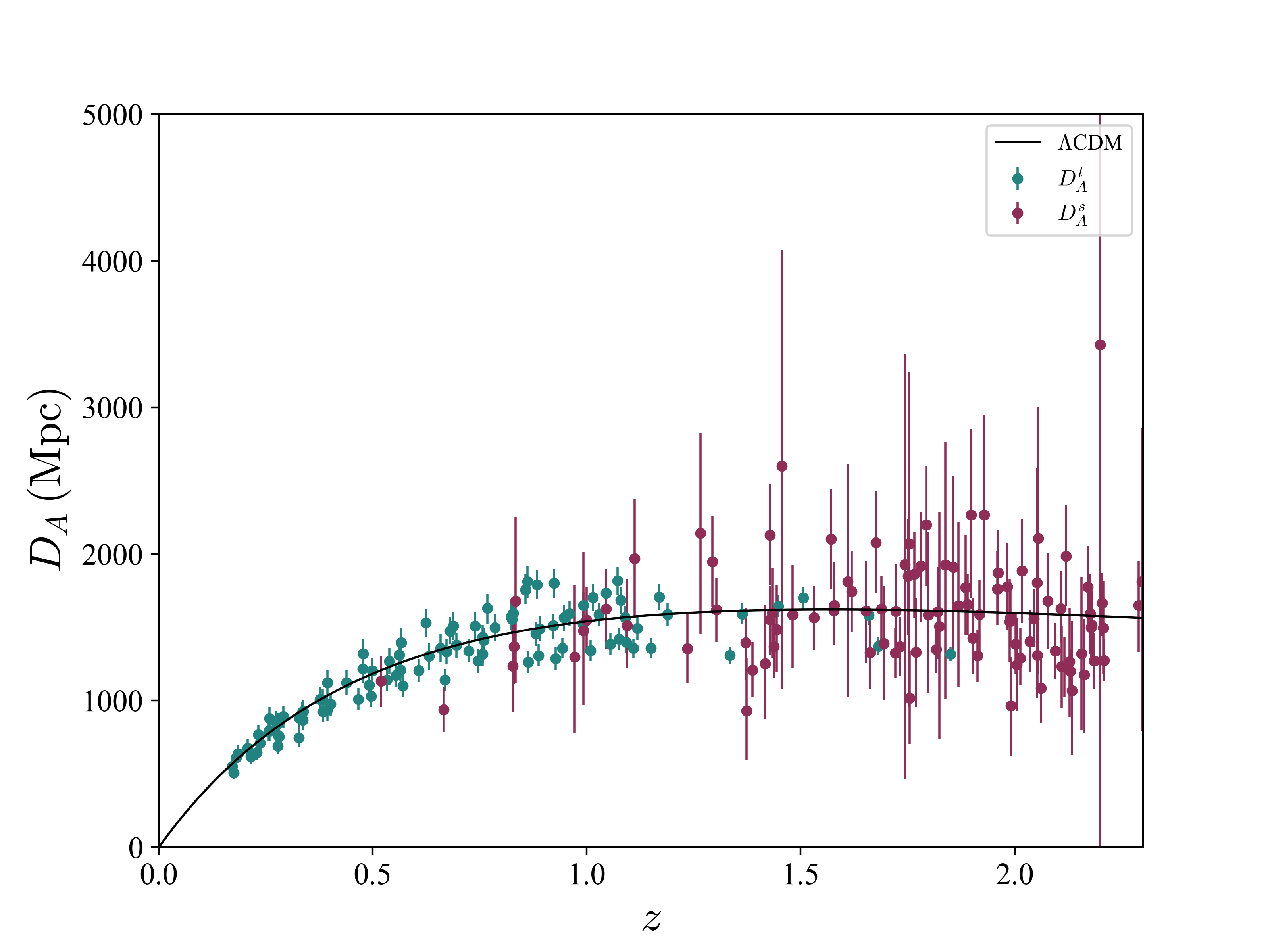}
 \caption{The $N=100$ mock SGL syetems in an arbitrary realization of simulation. Left panel: the redshift distribution of $z_s$ and $z_l$; Right panel: the angular diameter distances $D_A$ and their $1\sigma$ uncertainties. The black line is the fiducial $\Lambda$CDM curve.}\label{fig:sim_data}
\end{figure*}

We repeat the above steps 100 times to generate 100 SGL systems, resulting in a simulated sample including 200 angular diameter distances, i.e. 100 $D_A^l$ and 100 $D_A^s$. The redshift distributions of $z_s$ and $z_l$ in an arbitrary realization of simulation are plotted in the left penal of Figure \ref{fig:sim_data}, and the angular diameter distances are presented in the right panel of Figure \ref{fig:sim_data}.

\begin{figure*}[htbp]
 \centering
 \includegraphics[width=0.48\textwidth]{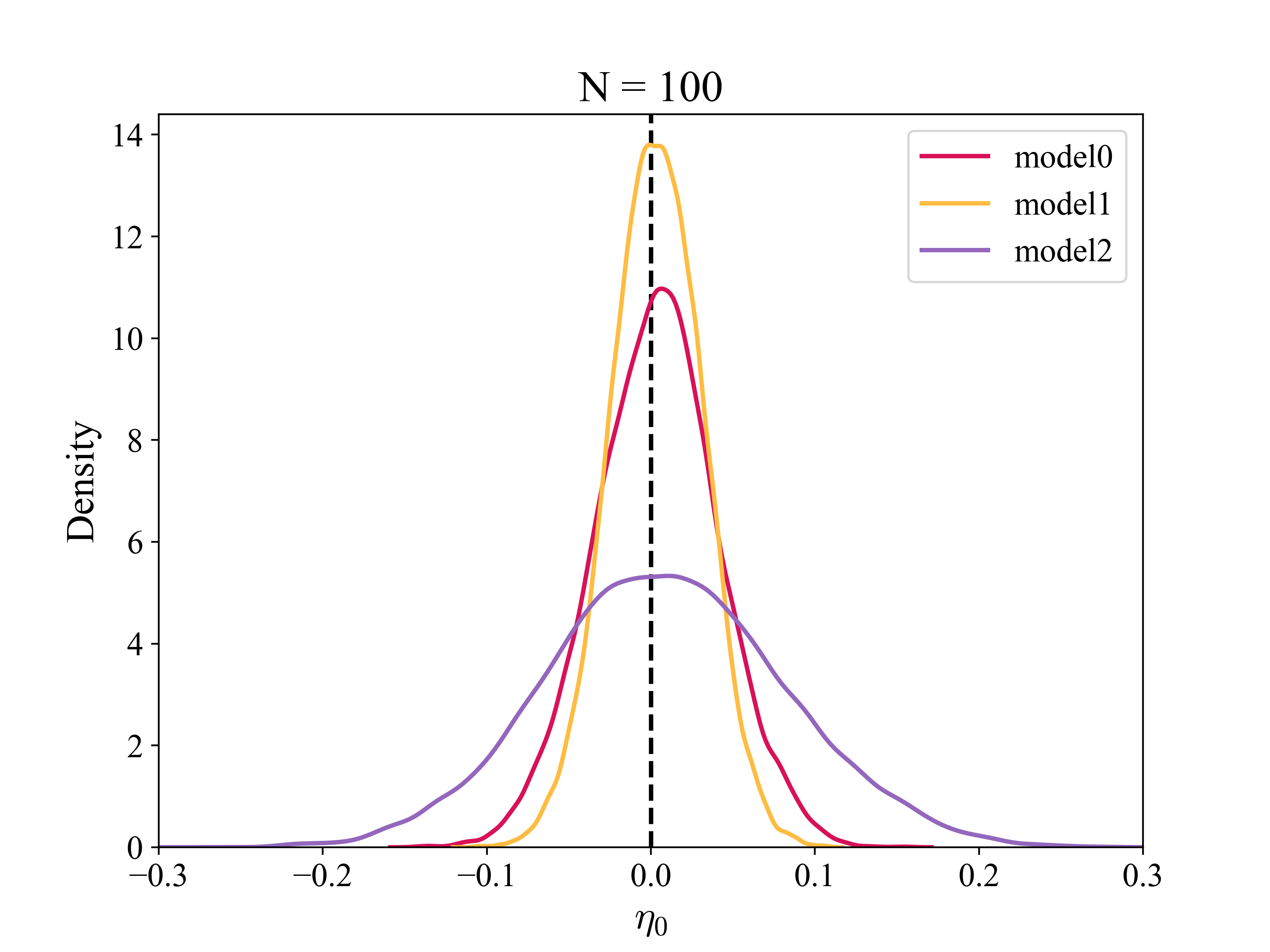}
 \includegraphics[width=0.48\textwidth]{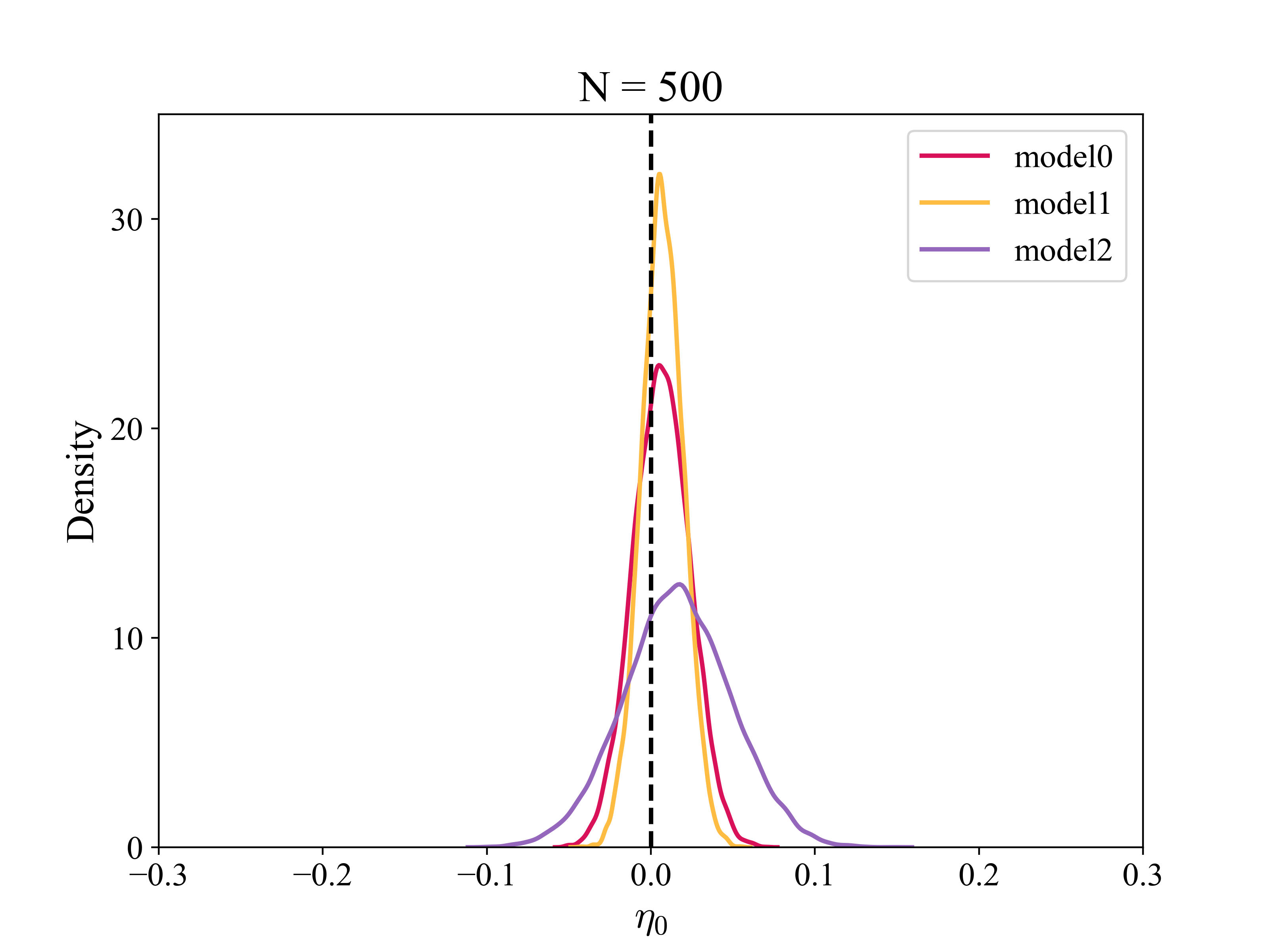}
 \caption{The posterior PDFs of $\eta_0$ in three parameterized models, constrained from  simulated SGL systems. Left panel: $N=100$; Right panel: $N=500$.} \label{fig:eta0_sim}
\end{figure*}

Combining the simulated angular diameter distance sample and the luminosity distance from SNe Ia, we constrain the DDR violation parameter $\eta_0$ in three parameterized models in the same manner as is described in Section {\ref{sec:DDR}}. The results are presented in Table \ref{tab:parameters_sim} and Figure \ref{fig:eta0_sim}. With 100 simulated SGL systems based on future observation accuracy, the parameter $\eta_0$ are strictly constrained in three models, with precision at the $10^{-2}$ level. Additionally, we also simulate 500 SGL systems to further constrain $\eta_0$, with the results again shown in Table \ref{tab:parameters_sim} and Figure \ref{fig:eta0_sim}. These simulations indicate that increasing the number of SGL systems from $N=100$ to $N=500$ reduces the uncertainty by an additional factor of about two, consistent with the usual $\sigma\propto N^{-1/2}$ relation.

\begin{table}[htbp]
\centering
\caption{\small{The best-fitting parameter $\eta_0$ in three parameterized models, constrained from simulated SGL systems.}} \label{tab:parameters_sim}
\arrayrulewidth=1.0pt
\renewcommand{\arraystretch}{1.3}
{\begin{tabular}{cccc} 
\hline\hline 
    $N$ & Model 0& Model 1 & Model 2\\\hline
100 & $0.004_{-0.038}^{+0.036}$ & $0.004_{-0.028}^{+0.028}$ &$0.008_{-0.071}^{+0.074}$ \\
500 & $0.006_{-0.017}^{+0.017}$ & $0.007_{-0.012}^{+0.013}$ &$0.016_{-0.032}^{+0.033}$ \\
\hline
\end{tabular}}
\end{table}

\section{Discussion and conclusions}\label{sec:conclusions}

The cosmic DDR that connects the angular diameter distance and the luminosity distance is a fundamental relation in cosmology. Any deviation from the DDR suggests the presence of new physics. Consequently, numerous model-independent methods and datasets have been employed to test the DDR. Advancements in observational techniques and data analysis methods are pivotal in reducing uncertainties and enhancing the accuracy of cosmological tests. Recently, the H0LiCOW collaboration constrained the Hubble constant $H_0$ to a precision of 2.4$\%$ using six gravitationally lensed quasars with precisely measured time-delay distances. The time-delay distance is related to the angular diameter distance, offering a promising avenue for high-precision DDR tests. In this study, we test the DDR by utilizing the angular diameter distances from the H0LiCOW collaboration's SGL sample, which provides precise time-delay distances and angular diameter distances to the lens for four SGL systems, along with luminosity distances derived from the latest Pantheon+ compilation of SNe Ia.

To minimize the statistical error arising from the redshift matching of SNe Ia with the SGL systems, we employed the GP method to reconstruct the $D_L-z$ relation from the Pantheon+ dataset, thereby determining the luminosity distances of the SGL systems. By combining the angular diameter distance from SGL systems with the GP reconstruction of luminosity distence from SNe Ia, we constrained the parameter $\eta_0$, which quantifies potential deviations from the standard DDR, across three parameterized models. The parameter $\eta_0$ was constrained to $\eta_0=-0.078_{-0.196}^{+0.200}$ in Model 0, $\eta_0=-0.081_{-0.186}^{+0.195}$ in Model 1, and $\eta_0=-0.166_{-0.405}^{+0.433}$ in Model 2, thus confirming the validity of the DDR at the 1$\sigma$ confidence level. Furthermore, we investigated the potential precision constraints based on the future observational accuracy using Monte Carlo simulations. With 100 simulated SGL systems, the precision of $\eta_0$ could reach the $10^{-2}$ level across all three parameterized models.

Previous studies have tested the DDR by combing different samples. Using the Union2 sample of SNe Ia, Li et al. \cite{Li:2011exa} parameterized the DDR with Model 1 and found that the DDR is valid at the 1$\sigma$ confidence level with a constraint of $\eta_0=-0.07^{+0.19}_{-0.19}$ from 18 cluster galaxies, but is violated at the 1$\sigma$ confidence level with a constraint of $\eta_0=-0.22^{+0.11}_{-0.11}$ from 38 cluster galaxies. Combining the Einstein angular measurement of 161 SGL systems with JLA sample of SNe Ia, Liao et al. \cite{Liao:2015uzb} constrained the parameter $\eta_0$ of Model 1 to $\eta_0=-0.005^{+0.351}_{-0.215}$. Similarly, combining 205 SGL systems and Pantheon sample, Zhou et al. \cite{Zhou:2020moc} constrained $\eta_0=0.047^{+0.190}_{-0.151}$, verifying the DDR at the 1$\sigma$ confidence level. In our work, by using only four SGL systems with measured time-delay distance, we achieved a constraint on the DDR parameter with accuracy comparable to the aforementioned results. Furthermore, our simulation results suggest that the upcoming observational advancements from the LSST survey's lensed quasars will significantly enhance the precision of DDR testing.

In the previous work of Liao et al. \cite{Liao:2015uzb}, the distance ratio derived from Einstein ring is used to constrain DDR. The Einstein ring depends on the ratio of distance, $R_A=D_A^{ls}/D_A^s$. Although the available sample of SGL systems is large, the ratio of distance may cancel some important information, thus may introduce large uncertainty on testing DDR. While in our work, the degeneracy between $D_A^{ls}$ and $D_A^s$ is broken by the observed time delay, then the angular diameter distances from the observer to the lens and to the source are obtained simultaneously. The direct comparison of the angular diameter distance and the luminosity distance provides a more precise way to test DDR than using the distance radio. For example, with only four SGL systems, the DDR violation parameter $\eta_0$ can be constrained at the precision of $0.2$ in Model 1, which is comparable to the precision of 60 SGL systems obtained in Ref.\cite{Liao:2015uzb}. However, our method is more susceptible to systematic errors from the absolute calibration of time-delay distances, such as uncertainties in the lens mass distribution, the line-of-sight structure, and the external convergence effects. In addition, at present a majority of SGL systems have no time delay measurement, so the available data sample is small. We expect that the LSST survey will provide more time-delayed SGL systems, hence can give a tight constraint on DDR in the near future.

\vspace{5mm}
\centerline{\rule{80mm}{0.5pt}}
\vspace{2mm}

\bibliographystyle{cpc}
\bibliography{reference}

\begin{thebibliography}{10}
\providecommand{\url}[1]{\texttt{#1}}
\providecommand{\urlprefix}{URL }
\providecommand{\eprint}[2][]{\url{#2}}

\bibitem{etherington1933lx}
I.~Etherington.
\newblock The London, Edinburgh, and Dublin Philosophical Magazine and Journal
  of Science, \textbf{15}~(100): 761--773 (1933).

\bibitem{Bassett:2003vu}
B.~A. Bassett and M.~Kunz.
\newblock Phys. Rev. D, \textbf{69}: 101305 (2004).

\bibitem{Corasaniti:2006cv}
P.~S. Corasaniti.
\newblock Mon. Not. Roy. Astron. Soc., \textbf{372}: 191 (2006).

\bibitem{Ellis:2013cu}
G.~F.~R. Ellis, R.~Poltis, J.-P. Uzan \emph{et~al.}
\newblock Phys. Rev. D, \textbf{87}~(10): 103530 (2013).

\bibitem{Holanda:2012ia}
R.~F.~L. Holanda, J.~C. Carvalho, and J.~S. Alcaniz.
\newblock JCAP, \textbf{04}: 027 (2013).

\bibitem{Cao:2016ggq}
S.~Cao, M.~Biesiada, X.~Zheng \emph{et~al.}
\newblock Mon. Not. Roy. Astron. Soc., \textbf{457}~(1): 281--287 (2016).

\bibitem{Holanda:2011hh}
R.~F.~L. Holanda, J.~A.~S. Lima, and M.~B. Ribeiro.
\newblock Astron. Astrophys., \textbf{538}: A131 (2012).

\bibitem{Kumar:2020ole}
D.~Kumar, D.~Jain, S.~Mahajan \emph{et~al.}
\newblock Phys. Rev. D, \textbf{103}~(6): 063511 (2021).

\bibitem{Liao:2020zko}
K.~Liao, A.~Shafieloo, R.~E. Keeley \emph{et~al.}
\newblock Astrophys. J. Lett., \textbf{895}~(2): L29 (2020).

\bibitem{DeBernardis:2006ii}
F.~De~Bernardis, E.~Giusarma, and A.~Melchiorri.
\newblock Int. J. Mod. Phys. D, \textbf{15}: 759--766 (2006).

\bibitem{Holanda:2010ay}
R.~F.~L. Holanda, J.~A.~S. Lima, and M.~B. Ribeiro.
\newblock Astron. Astrophys., \textbf{528}: L14 (2011).

\bibitem{Uzan:2004my}
J.-P. Uzan, N.~Aghanim, and Y.~Mellier.
\newblock Phys. Rev. D, \textbf{70}: 083533 (2004).

\bibitem{Avgoustidis:2010ju}
A.~Avgoustidis, C.~Burrage, J.~Redondo \emph{et~al.}
\newblock JCAP, \textbf{10}: 024 (2010).

\bibitem{Hu:2018yah}
J.~Hu and F.~Y. Wang.
\newblock Mon. Not. Roy. Astron. Soc., \textbf{477}~(4): 5064--5071 (2018).

\bibitem{Meng:2011nt}
X.-L. Meng, T.-J. Zhang, and H.~Zhan.
\newblock Astrophys. J., \textbf{745}: 98 (2012).

\bibitem{Li:2011exa}
Z.~Li, P.~Wu, and H.~W. Yu.
\newblock Astrophys. J. Lett., \textbf{729}: L14 (2011).

\bibitem{Santos-da-Costa:2015kmv}
S.~Santos-da Costa, V.~C. Busti, and R.~F.~L. Holanda.
\newblock JCAP, \textbf{10}: 061 (2015).

\bibitem{Liao:2015uzb}
K.~Liao, Z.~Li, S.~Cao \emph{et~al.}
\newblock Astrophys. J., \textbf{822}~(2): 74 (2016).

\bibitem{Zhou:2020moc}
C.~Zhou, J.~Hu, M.~Li \emph{et~al.}
\newblock Astrophys. J., \textbf{909}~(2): 118 (2021).

\bibitem{Kumar:2021djt}
D.~Kumar, A.~Rana, D.~Jain \emph{et~al.}
\newblock JCAP, \textbf{01}~(01): 053 (2022).

\bibitem{He:2022phb}
Y.~He, Y.~Pan, D.-P. Shi \emph{et~al.}
\newblock Chin. J. Phys., \textbf{78}: 297--307 (2022).

\bibitem{Tang:2022ykd}
L.~Tang, H.-N. Lin, and L.~Liu.
\newblock Chin. Phys. C, \textbf{47}~(1): 015101 (2023).

\bibitem{Tonghua:2023hdz}
L.~Tonghua, C.~Shuo, M.~Shuai \emph{et~al.}
\newblock Phys. Lett. B, \textbf{838}: 137687 (2023).

\bibitem{Lin:2018qal}
H.-N. Lin, M.-H. Li, and X.~Li.
\newblock Mon. Not. Roy. Astron. Soc., \textbf{480}~(3): 3117--3122 (2018).

\bibitem{Liao:2019xug}
K.~Liao.
\newblock Astrophys. J., \textbf{885}: 70 (2019).

\bibitem{Lin:2019mrl}
H.-N. Lin and X.~Li.
\newblock Chin. Phys. C, \textbf{44}~(7): 075101 (2020).

\bibitem{Lin:2020vqj}
H.-N. Lin, X.~Li, and L.~Tang.
\newblock Chin. Phys. C, \textbf{45}~(1): 015109 (2021).

\bibitem{Arjona:2020axn}
R.~Arjona, H.-N. Lin, S.~Nesseris \emph{et~al.}
\newblock Phys. Rev. D, \textbf{103}~(10): 103513 (2021).

\bibitem{Renzi:2020bvl}
F.~Renzi, N.~B. Hogg, M.~Martinelli \emph{et~al.}
\newblock Phys. Dark Univ., \textbf{32}: 100824 (2021).

\bibitem{Huang:2024zvk}
S.-J. Huang, E.-K. Li, J.-d. Zhang \emph{et~al.}
\newblock arXiv:2402.17349 (2024).

\bibitem{Scolnic:2021amr}
D.~Scolnic \emph{et~al.}
\newblock Astrophys. J., \textbf{938}~(2): 113 (2022).

\bibitem{Brout:2021mpj}
D.~Brout \emph{et~al.}
\newblock Astrophys. J., \textbf{938}~(2): 111 (2022).

\bibitem{Suyu:2009by}
S.~H. Suyu, P.~J. Marshall, M.~W. Auger \emph{et~al.}
\newblock Astrophys. J., \textbf{711}: 201--221 (2010).

\bibitem{Suyu:2013kha}
S.~H. Suyu \emph{et~al.}
\newblock Astrophys. J. Lett., \textbf{788}: L35 (2014).

\bibitem{Suyu:2016qxx}
S.~H. Suyu \emph{et~al.}
\newblock Mon. Not. Roy. Astron. Soc., \textbf{468}~(3): 2590--2604 (2017).

\bibitem{Wong:2016dpo}
K.~C. Wong \emph{et~al.}
\newblock Mon. Not. Roy. Astron. Soc., \textbf{465}~(4): 4895--4913 (2017).

\bibitem{Jee:2019hah}
I.~Jee, S.~Suyu, E.~Komatsu \emph{et~al.}
\newblock Science, \textbf{365}~(6458): 1134--1138 (2019).

\bibitem{Chen:2019ejq}
G.~C.~F. Chen \emph{et~al.}
\newblock Mon. Not. Roy. Astron. Soc., \textbf{490}~(2): 1743--1773 (2019).

\bibitem{Birrer:2018vtm}
S.~Birrer \emph{et~al.}
\newblock Mon. Not. Roy. Astron. Soc., \textbf{484}: 4726 (2019).

\bibitem{Rusu:2019xrq}
C.~E. Rusu \emph{et~al.}
\newblock Mon. Not. Roy. Astron. Soc., \textbf{498}~(1): 1440--1468 (2020).

\bibitem{Wong:2019kwg}
K.~C. Wong \emph{et~al.}
\newblock Mon. Not. Roy. Astron. Soc., \textbf{498}~(1): 1420--1439 (2020).

\bibitem{Mollerach:2002}
S.~{Mollerach} and E.~{Roulet}.
\newblock \emph{{Gravitational Lensing and Microlensing}},  (World Scientific,
  Singapore2002).

\bibitem{Planck:2018vyg}
N.~Aghanim \emph{et~al.} (Planck).
\newblock Astron. Astrophys., \textbf{641}: A6 (2020).
\newblock [Erratum: Astron.Astrophys. 652, C4 (2021)].

\bibitem{Bartelmann:1999yn}
M.~Bartelmann and P.~Schneider.
\newblock Phys. Rept., \textbf{340}: 291--472 (2001).

\bibitem{Eigenbrod:2005ie}
A.~Eigenbrod, F.~Courbin, C.~Vuissoz \emph{et~al.}
\newblock Astron. Astrophys., \textbf{436}: 25 (2005).

\bibitem{Bonvin:2018dcc}
V.~Bonvin \emph{et~al.}
\newblock Astron. Astrophys., \textbf{616}: A183 (2018).

\bibitem{Tripp:1997wt}
R.~Tripp.
\newblock Astron. Astrophys., \textbf{331}: 815--820 (1998).

\bibitem{Kessler:2016uwi}
R.~Kessler and D.~Scolnic.
\newblock Astrophys. J., \textbf{836}~(1): 56 (2017).

\bibitem{scikit-learn}
F.~Pedregosa, G.~Varoquaux, A.~Gramfort \emph{et~al.}
\newblock Journal of Machine Learning Research, \textbf{12}: 2825--2830 (2011).

\bibitem{Brout:2022vxf}
D.~Brout \emph{et~al.}
\newblock Astrophys. J., \textbf{938}~(2): 110 (2022).

\bibitem{Oguri:2010ns}
M.~Oguri and P.~J. Marshall.
\newblock Mon. Not. Roy. Astron. Soc., \textbf{405}: 2579--2593 (2010).

\bibitem{Liao:2017ioi}
K.~Liao, X.-L. Fan, X.-H. Ding \emph{et~al.}
\newblock Nature Commun., \textbf{8}~(1): 1148 (2017).
\newblock [Erratum: Nature Commun. 8, 2136 (2017)].

\bibitem{Cao:2019kgn}
S.~Cao, J.~Qi, Z.~Cao \emph{et~al.}
\newblock Sci. Rep., \textbf{9}~(1): 11608 (2019).

\bibitem{Biesiada:2014kwa}
M.~Biesiada, X.~Ding, A.~Piorkowska \emph{et~al.}
\newblock JCAP, \textbf{10}: 080 (2014).

\end{thebibliography}

\end{document}